\documentclass[prb,preprint,titlepage,amsmath,amssymb,superscriptaddress,longbibliography]{revtex4-2}

\usepackage{graphicx}
\usepackage{dcolumn}
\usepackage{bm}

\usepackage{soul}

\newcommand{\RN}[1]{%
	\textup{\uppercase\expandafter{\romannumeral#1}}%
}

\usepackage[inner=1.5cm,outer=1.5cm,bottom=2cm,tmargin=1cm]{geometry}

\usepackage{mathtools}

\usepackage{outlines}
\usepackage{fancyvrb}
\usepackage{amsmath,nccmath}
\usepackage{hyperref}
\usepackage{braket}
\usepackage{bbold}

\usepackage{amssymb}

\usepackage[usenames, dvipsnames]{color}

\usepackage{float}

\usepackage{rotating}

\usepackage{tensor}

\usepackage{dsfont}
\usepackage{physics}

\usepackage{diagbox}
\usepackage{stackengine}
\usepackage[normalem]{ulem}
\usepackage{cancel}

\begin{document}
\title{Electro-optic sampling of the electric-field operator for ultrabroadband pulses of Gaussian quantum light}
\author{Geehyun Yang}

\affiliation{Department of Physics, KAIST, Daejeon, Republic of Korea}

\author{Sandeep Sharma}

\affiliation{Department of Physics, KAIST, Daejeon, Republic of Korea}

\author{Andrey S. Moskalenko}

\email{moskalenko@kaist.ac.kr }

\affiliation{Department of Physics, KAIST, Daejeon, Republic of Korea}

\begin{abstract}

Quantum light pulses (QLPs) can be described by spatio-temporal modes, each of which is associated with a quantum state.
In the mid-infrared spectral range, electro-optic sampling (EOS) provides a means to characterize quantum fluctuations in the electric field of such light pulses.
Here, we present a protocol based on the two-port EOS technique that enables the complete characterization of multimode Gaussian quantum light, demonstrating robustness to both the shot noise and cascading effects.
We validate this approach theoretically by reconstructing a multimode squeezed state of light generated in a thin nonlinear crystal driven by a single-cycle pulse.
Our findings establish the two-port EOS technique as a versatile tool for characterizing ultrafast multimode quantum light, thereby broadening the reach of quantum state tomography.
Potential applications include
the characterization of complex quantum structures, such as correlations and entanglement in light and matter. Further, extensions to study multimode non-Gaussian QLPs can be envisaged.

\end{abstract}

\maketitle

\section{Introduction}

Generation and detection of quantum light are essential for realizing efficient quantum technologies, including quantum sensing and metrology \cite{Caves1981,Xiao1987,Grangier1987,LIGOgravitationalwave} as well as quantum information processing \cite{Menicucci2006,Braunstein2005,Spring2013,Tillmann2013,Crespi2013,Spagnolo2014}.
Among the various methods for characterizing the quantum nature of light, phase-space tomography stands out due to its ability to fully visualize quantum states.
Towards this, conventional approaches, such as homodyne and heterodyne detection are frequently employed to reconstruct phase-space distributions like the Wigner and Husimi-Q functions, and subsequently identify the quantum state of light.
In the mid-infrared (MIR) frequency range, a novel approach, subcycle tomography based on an ultrafast version of the conventional homodyne detection or, alternatively, on a generalization of the electro-optic sampling (EOS) technique \cite{Optica_review2025} to the quantum domain \cite{Riek2015,MoskalenkoParaxial,THz_Roadmap2023,Optica_review2025}, allowing to circumvent the problem of the quantum efficiency of MIR photodetectors, has been proposed recently in some variations \cite{Yang2023,Onoe2023,Emanuel2024}.
While subcycle tomography has proven effective for observing temporally localized properties of the QLP, it falls short in capturing the correlation properties contained in its multimode structure, preventing the full characterization of such light.
In this work, we introduce a two-port EOS technique that allows for the reconstruction of the full multimode structure of ultrashort Gaussian QLPs.
We demonstrate such reconstruction by considering an example of an ultrabroadband MIR squeezed vacuum state generated by a single-cycle pulse in a thin $\chi^{(2)}$ generation crystal.

Direct detection of the MIR vacuum fluctuations and correlations through the EOS has been shown
both theoretically \cite{MoskalenkoParaxial,Lindel2024,Lindel2020} and experimentally \cite{RiekSubcycle,Benea2019,Settembrini2022}.
To extend the approach based on the usual EOS setup towards subcycle tomography it happened to be necessary to rely on highly non-trivial spectral filtering of the detected photons in the near infrared (NIR), to which the EOS upconverts the MIR signal photons for counting by balanced detectors \cite{Gallot1999}.
Despite its ability to fully characterize QLPs locally in the general case, as well as also globally in the case of pure Gaussian states reducible to a single temporally localized mode, subcycle tomography traces out the multimode structure outside of the space of the localized detection mode, resulting in an incoherent mixture of contributions from each mode \cite{Yang2023}, which are not possible to disentangle.
We show that by upgrading the approach to the two-port EOS technique \cite{Benea2019},
the feat of determining the whole multimode structure of an ultrashort Gaussian QLP can be accomplished, capturing correlations between different localized modes of an arbitrary complete basis and, thus, also between different time bins.
Notably, this approach does not rely on spectral filtering, avoiding the related complications and suggesting itself as a new technique for the complete characterization of Gaussian QLPs in the time domain, advancing the classical EOS to the quantum regime in this respect.

This paper is organized as follows: In Sec.~\ref{sec:One-port EOS}, we present the single-port EOS setup and discuss the detection of the EOS signal and the corresponding mode operator.
Next, we provide an analytical expression for the mode operator within the framework of second-order perturbation theory and explore the detection of the MIR and NIR components of the mode operator via the estimation of the corresponding signal variance.
In Sec.~\ref{sec:two-port EOS}, we present
the two-port EOS scheme, which enables the measurement of temporal correlations in ultrafast MIR quantum light.
In Sec.~\ref{sec:Reconstruction of Gaussian quantum light pulse}, we describe the complete reconstruction of a multimode Gaussian QLP based on the observed temporal correlations.
In Sec.~\ref{sec:Discussion}, we provide a short discussion on this reconstruction.
Finally, in Sec.\ref{sec:conclusion}, we summarize our work.

\section{Single-port electro-optic sampling}\label{sec:One-port EOS}

\begin{figure}[t]\includegraphics[width=18cm]{"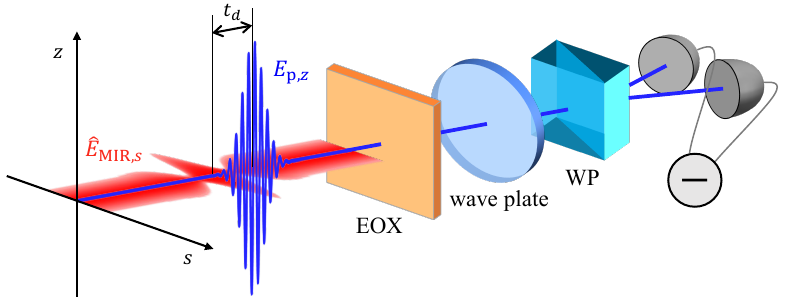"}
	\caption{
		\label{Fig:One port EOS}
		 Schematic of the single-port EOS. The incoming MIR quantum light $\hat{E}_{\mathrm{MIR},s}$ (red band) interacts in an electro-optic crystal (EOX) with a co-propagating NIR probe pulse $E_{\mathrm{p},z}$ (blue) centered at a variable time delay $t_\mathrm{d}$. The interaction induces a change
in the polarization state of the probe, which is analyzed by the following adjustable ellipsometer consisting of a variable phase-shift wave plate, a Wollaston prism (WP), and two balanced photodetectors that receive probe photons with mutually perpendicular polarization. The orientation of the fast axis of the wave plate is adjusted through its rotation around the propagation axis, ensuring equal mean probe photon numbers detected by each photodetector when purely quantum (or just vacuum) MIR field is analyzed \cite{Sulzer2020,Kizmann2022,Unruh-DeWitt}.}
		\label{EOS}
\end{figure}

We consider the conventional single-port EOS setup \cite{Optica_review2025} as shown in Fig.~\ref{Fig:One port EOS}. An intense NIR probe pulse $E_{\mathrm{p},z}(t)$, polarized along the $z$-direction, interacts with a co-propagating
MIR electric field $\hat{E}_{\mathrm{MIR},s}(t)$, polarized along the $s$-direction, inside an electro-optic crystal (EOX). Interaction between both of these fields gives rise to a weak NIR field $\hat{E}'_{\mathrm{p},s}(t)$, with polarization along the $s$-direction, via a three-wave mixing process \cite{Gallot1999,MoskalenkoParaxial,Optica_review2025,Kizmann2022Theory,Yang2023}.
The modified NIR field then passes through a variable wave plate, which changes its polarization state by inducing a phase shift $\theta$ for the component along the fast axis of the wave plate, whereas the balancing of the final signal is assured as described in the caption to
Fig.~\ref{Fig:One port EOS}. The following Wollaston prism (WP)
separates two orthogonal polarization components of the modified NIR probe, with
the number of photons in each component
counted by the corresponding photodector.
Finally, the difference in those photon numbers
leads to the resulting electro-optic signal given by
 \begin{eqnarray}\label{eq:eos-signal}
 	\hat{S}(\theta;t_\mathrm{d})=C\int_0^\infty \mathrm{d}\omega \, \frac{1}{\hbar \omega} \left[P(\theta) \tilde{E}^*_{\mathrm{p},z}(\omega) \hat{E}'_{\mathrm{p},s}(\omega) +\text{H.c.}\right],
 \end{eqnarray}
where $C=4\pi\varepsilon_0 c_0 nA$ with $c_0$ being the speed of light in the vacuum, $\varepsilon_0$ is the vacuum permittivity, $n$ is refractive index (RI) at the frequencies of the probe,
$A$ is the effective transverse area determined by probe beam waist, and for the probe pulse we implicitly assume its delay $t_\mathrm{d}$ with respect to the sampled MIR field. For a field $E(t)$, $\tilde{E}(\omega)$ denotes the Fourier-transform of $\tilde{E}(t)\equiv E(t-t_\mathrm{d})$ including delay $t_\mathrm{d}$, to differentiate from $E(\omega)$ being the Fourier transform of $E(t)$.
The factor $P(\theta)=\sqrt{-\cos\theta}+i\sqrt{2}\cos(\theta/2)$ reflects the resulting action of the variable phase shifter, whereas $\theta$
is bounded to the interval $\frac{\pi}{2} \leq \theta \leq \frac{3\pi}{2}$ \cite{Kizmann2022,Kizmann2022Theory,Yang2023}.
Further, the detected electro-optic signal can also be expressed as
 \begin{eqnarray}\label{eq:field_oper}
 	\hat{S}(\theta;t_\mathrm{d})=\sqrt{N}\left[P(\theta)\hat{b}(t_\mathrm{d})+\text{H.c.}\right],
 \end{eqnarray}
where $N=C\int_0^\infty \mathrm{d}\omega \, \frac{1}{\hbar \omega} \lvert  E_{\mathrm{p},z}(\omega) \rvert^2$ is the mean photon number in the probe, reflecting also the shot-noise level of the probe, and
 \begin{eqnarray}\label{eq:field_oper_2}
    \hat{b}(t_\mathrm{d})=\int_0^\infty \mathrm{d} \omega \, h(\omega;t_\mathrm{d}) \hat{b}(\omega),\quad
 	h(\omega;t_\mathrm{d})=i\sqrt{\frac{C}{\hbar N}} \frac{E_{\mathrm{p},z}^*(\omega)e^{-i\omega t_\mathrm{d}}}{\sqrt{\omega}}.
 \end{eqnarray}
Here $\hat{b}(\omega)$ is the annihilation operator of the probe.

Next, we employ a perturbative approach to treat the three-wave mixing process in the EOX, resulting an analytical expression for
$\hat{b}(t_\mathrm{d})$
in terms of the field operators corresponding to $\hat{E}_{\mathrm{MIR},s}(t)$ and $\hat{E}_{\mathrm{p},s}(t)$.
This nonlinear process is governed by the interaction Hamiltonian
 \begin{eqnarray}\label{eq:three_wave}
 	\hat{H}_{\mathrm{int}}(t)=\epsilon_0 d \int_V \mathrm{d}^3 r \, \tilde{E}^{(+)}_{\mathrm{p},z}(\vec{r},t) \left[ \hat{E}^{(+)}_{\mathrm{MIR},s}(\vec{r},t) +\hat{E}^{(-)}_{\mathrm{MIR},s}(\vec{r},t) \right]\hat{E}^{(-)}_{\mathrm{p},s}(\vec{r},t) +\text{H.c.},
 \end{eqnarray}
where $V$ denotes volume of the EOX, $d$ is the effective nonlinear susceptibility of the EOX, and the superscripts $(+)$ and $(-)$ correspond to the positive and negative frequency parts of the electric field, respectively.
The electric field components can be expressed in terms of the field operator $\hat{a}(\omega;t)$ as $\hat{E}^{(+)}(\vec{r},t)=i \int_0^\infty \mathrm{d}\omega \, \sqrt{\frac{\hbar \omega}{C}} \hat{a}(\omega;t) e^{i(k(\omega) r_{\parallel}-\omega t)}$ and $\hat{E}^{(-)}(\vec{r},t)=\big(\hat{E}^{(+)}(\vec{r},t)\big)^\dagger$, where $k(\omega)$ is the wave vector and $r_{\parallel}$ denotes the position component parallel to the propagation direction.
Subsequently, the Fourier components of the field can be written as $\hat{E}(\omega;t)=i\, \text{sgn}(\omega)  \sqrt{\frac{\hbar \lvert \omega \rvert}{C}} \hat{a}(\omega;t)$.
For the input electric field and its corresponding field operator, we omit the explicit time dependence and express them as $\hat{E}(\omega)=\hat{E}(\omega;-\infty)$ and $\hat{a}(\omega)=\hat{a}(\omega;-\infty)$, respectively.
We analyze the three-wave mixing process within the framework of second-order perturbation theory and express the $s$-polarized NIR field as $\hat{E}'_{\mathrm{p},s}(\omega)=\hat{E}_{\mathrm{p},s}(\omega)+\hat{E}^{(1)}_{\mathrm{p},s}(\omega)+\hat{E}^{(2)}_{\mathrm{p},s}(\omega)$.

The first-order perturbative term $\hat{E}^{(1)}_{\mathrm{p},s}(\omega)$ can be derived using the following equation
 \begin{eqnarray}\label{eq:pert_1}
 	\hat{E}_{\mathrm{p},s}^{(1)}(\omega)=\frac{i}{\hbar}\int_{-\infty}^{\infty} \mathrm{d}t \, \big[\hat{H}_{\mathrm{int}}(t),\hat{E}_{\mathrm{p},s}(\omega) \big].
 \end{eqnarray}
Evaluation of the above equation by making use of Eq.~\eqref{eq:three_wave}, leads to
 \begin{eqnarray}\label{eq:first order NIR field}
 	\hat{E}_{\mathrm{p},s}^{(1)}(\omega)=-\frac{idl\omega}{2c_0n}\int \mathrm{d}\Omega \, \hat{E}_{\mathrm{MIR},s}(\Omega) \tilde{E}_{\mathrm{p},z}(\omega-\Omega) F_1(\Omega),
 \end{eqnarray}
where
\begin{eqnarray}\label{eq:F1}
 	 F_1(\Omega)=\mathrm{sinc}\bigg[\frac{l\Omega}{2c_0}(n(\Omega)-n_\mathrm{g})\bigg],
\end{eqnarray}
$l$ is the length of the EOX, $n(\Omega)$ is the RI at the frequencies of MIR field, and $n_\mathrm{g}$ is the group RI of the probe pulse \cite{MoskalenkoParaxial,Guedes2023}.
From here on, we denote $\Omega$ as the MIR frequency and $\omega$ as the NIR frequency. If the integral limits are not indicated, this means that the integration goes over the whole relevant frequency range, including positive and negative frequencies.
The second-order perturbative term $\hat{E}^{(2)}_{\mathrm{p},s}(\omega)$ can be derived from
 \begin{eqnarray}\label{eq:pert_2}
 	\hat{E}^{(2)}_{\mathrm{p},s}(\omega)=-\frac{1}{\hbar^2}\int_{-\infty}^{\infty} \mathrm{d}t \int_{0}^{\infty} \mathrm{d}t' \big[ \hat{H}_{\mathrm{int}}(t), [\hat{H}_{\mathrm{int}}(t-t'),\hat{E}_{\mathrm{p},s}(\omega) ] \big] .
 \end{eqnarray}
Relying on Eq.~\eqref{eq:three_wave}, we obtain
  \begin{eqnarray}\label{eq:second order NIR field}
 	\hat{E}_{\mathrm{p},s}^{(2)}(\omega)=-\frac{2\pi d^2\omega l^2}{(4\pi c_0 n)^2} \int\mathrm{d}\omega' F_2(\omega,\omega')e^{i(\omega-\omega')t_\mathrm{d}} \hat{E}_{\mathrm{p},s}(\omega'),
\end{eqnarray}
where
 \begin{eqnarray}\label{eq:F2}
 	F_2(\omega,\omega')=\int \mathrm{d}\Omega \int \mathrm{d} \Omega'  \, \frac{n\Omega}{n(\Omega)}  E_{\mathrm{p},z}(\omega-\Omega') E^*_{\mathrm{p},z}(\omega'-\Omega')
 	 \text{sinc}^2 \Big[\frac{l}{2c_0}(n_\mathrm{g} \Omega'-n(\Omega)\Omega) \Big]g(\Omega'-\Omega),
 \end{eqnarray}
with
\begin{eqnarray}\label{eq:g}
 	g(x)=
 	\pi \delta(x)-\frac{i}{x}.
\end{eqnarray}
In the integrand of Eq.~\eqref{eq:F2}, the term $\pi\delta(\Omega'-\Omega)$ corresponds to cases where energy is conserved for each of the involved $\chi^{(2)}$ process separately, while the term $\frac{i}{\Omega'-\Omega}$ accounts for cases where energy is conserved only when both $\chi^{(2)}$ processes are considered together \cite{Kizmann2022,Kizmann2022Theory}.
We now make use of the relation $\hat{E}_{\mathrm{p},s}^{(1)}(\omega)=i\, \text{sgn}(\omega)  \sqrt{\frac{\hbar \lvert \omega \rvert}{C}} \hat{b}^{(1)}(\omega)$ as well as the similar expression for $\hat{E}_{\mathrm{MIR},s}(\Omega)$ and compare them with Eq.~\eqref{eq:first order NIR field}, to write the expression for the annihilation operator $\hat{b}^{(1)}(\omega)$ corresponding to the first-order perturbation of the NIR field as
\begin{eqnarray}\label{eq:first order NIR annihilation}
	\hat{b}^{(1)}(\omega)=-\frac{idl\sqrt{\omega}}{2c_0n}\int \mathrm{d}\Omega \,\text{sgn}(\Omega) \sqrt{\frac{n\,\lvert \Omega \rvert}{n(\Omega)}} \tilde{E}_{\mathrm{p},z}(\omega-\Omega) F_1(\Omega)  \hat{a}(\Omega),
\end{eqnarray}
where we use convention $\hat{a}(-\Omega)=\hat{a}^\dagger(\Omega)$ for $\Omega>0$.
Similarly, by using Eq.~\eqref{eq:second order NIR field}, the annihilation operator $\hat{b}^{(2)}(\omega)$ corresponding to the second-order perturbation of the NIR field can be expressed as
 \begin{eqnarray}\label{eq:second order NIR annihilation}
 		\hat{b}^{(2)}(\omega)=-\frac{2\pi d^2 l^2 \sqrt{\omega}}{(4\pi c_0 n)^2} \int\mathrm{d}\omega'  \, \text{sgn}(\omega') \sqrt{\lvert \omega' \rvert}
 F_2(\omega,\omega') e^{i(\omega-\omega')t_\mathrm{d}}
 \hat{a}(\omega').
 \end{eqnarray}
Also here $\hat{a}(-\omega)=\hat{a}^\dagger(\omega)$ for $\omega>0$.

Next, the detection mode operator $\hat{b}(t_\mathrm{d})$ corresponding to the electro-optic signal can be expressed in terms of the annihilation and creation operators of the MIR and NIR fields incoming to the EOX as
 \begin{eqnarray}\label{eq:field_oper_dec}
 	\begin{split}
 		\hat{b}(t_\mathrm{d})=\hat{\alpha}(t_\mathrm{d})+\hat{\beta}^\dagger (t_\mathrm{d})
 	\end{split}
 \end{eqnarray}
where
 \begin{eqnarray}\label{eq:field_oper_Thz}
 	\begin{split}
 		\hat{\alpha}(t_\mathrm{d})=&\hat{\alpha}_\mathrm{MIR}(t_\mathrm{d})+\hat{\alpha}_0(t_\mathrm{d})+\hat{\alpha}_\mathrm{NIR}(t_\mathrm{d}), \\
        \hat{\beta}(t_\mathrm{d})=&\hat{\beta}_\mathrm{MIR}(t_\mathrm{d})+\hat{\beta}_\mathrm{NIR}(t_\mathrm{d}),
 	\end{split}
 \end{eqnarray}
and $\hat{\alpha}_0(t_\mathrm{d})$ representing the zeroth-order detection mode operator.
The first-order perturbative term as in Eq.~\eqref{eq:first order NIR annihilation}
is determined by the field component in the MIR range.
Hence, by making use of Eq.~\eqref{eq:first order NIR annihilation}, we can express the MIR parts of the detection mode operator $\hat{b}(t_\mathrm{d})$ as
\begin{equation}\label{eq:THz detection mode}
	\begin{split}
		&\hat{\alpha}_{\mathrm{MIR}}(t_\mathrm{d})=\sqrt{\frac{C}{\hbar N}} \frac{dl}{2c_0n}\int_0^{\Omega_{\mathrm{max}}} \!\!\!\! \mathrm{d}\Omega \, \sqrt{\frac{n\Omega}{n(\Omega)}} G_-^*(\Omega) F_1(\Omega)   \hat{a}(\Omega)e^{-i\Omega t_\mathrm{d}},
		\\
		&\hat{\beta}_{\mathrm{MIR}}(t_\mathrm{d})=-\sqrt{\frac{C}{\hbar N}} \frac{dl}{2c_0n} \int_0^{\Omega_{\mathrm{max}}}\!\!\!\! \mathrm{d}\Omega \, \sqrt{\frac{n\Omega}{n(\Omega)}}  G_+(\Omega) F_1(\Omega) \hat{a}(\Omega) e^{-i\Omega t_\mathrm{d}},
	\end{split}
\end{equation}
where
 \begin{eqnarray}\label{eq:Gpm}
 	\begin{split}
 		G_{\pm}(\Omega)=\int_0^\infty \!\! \mathrm{d}\omega \, E_{\mathrm{p},z}(\omega) E^*_{\mathrm{p},z}(\omega\pm\Omega)
 	\end{split}
 \end{eqnarray}
and $\Omega_{\mathrm{max}}$ denotes maximum frequency of the MIR frequency range. Since the bandwidth of the probe is typically considerably smaller than its central frequency, we can actually simplify: $G_{+}(\Omega)=G_{-}^*(\Omega)\equiv G(\Omega)$.
Further, by making use of Eq.~\eqref{eq:second order NIR annihilation},
determined by the field component in the NIR range,
we can express the NIR parts of the detection mode operator $\hat{b}(t_\mathrm{d})$ as
\begin{equation}\label{eq:NIR alpha detection mode}
		\hat{\alpha}_{0}(t_\mathrm{d})+\hat{\alpha}_{\mathrm{NIR}}(t_\mathrm{d})= i \sqrt{\frac{C}{\hbar N}}  \int_0^\infty\!\!\!\!\mathrm{d}\omega \, \frac{E_{\mathrm{p},z}^*(\omega)}{\sqrt{\omega}}\left[\hat{a}(\omega)e^{-i\omega t_\mathrm{d}}- \frac{2\pi d^2 l^2 }{(4\pi c_0 n)^2}\!\! \int_0^{\infty}\!\!\!\!  \mathrm{d}\omega' \sqrt{\omega \omega' } F_2(\omega,\omega') \hat{a}(\omega') e^{-i\omega' t_\mathrm{d}}\right]\!,
\end{equation}
\begin{equation}\label{eq:NIR beta detection mode}
		\hat{\beta}_{\mathrm{NIR}}(t_\mathrm{d})=-i \sqrt{\frac{C}{\hbar N}} \frac{2\pi d^2 l^2 }{(4\pi c_0 n)^2}  \int_0^{\infty}\!\!\!\! \mathrm{d}\omega\,  E_{\mathrm{p},z}(\omega) \int_0^{\infty}\!\!\!\!  \mathrm{d}\omega'\, \sqrt{ \omega' }  F^*_2(\omega,-\omega')  \hat{a}(\omega')e^{-i\omega' t_\mathrm{d}} .
\end{equation}

We now investigate the detection mode $\hat{b}(t_\mathrm{d})$ in detail by analyzing its spectral amplitudes in both the MIR and NIR ranges under realizable experimental conditions.
\begin{figure}[t]\includegraphics[width=18cm]{"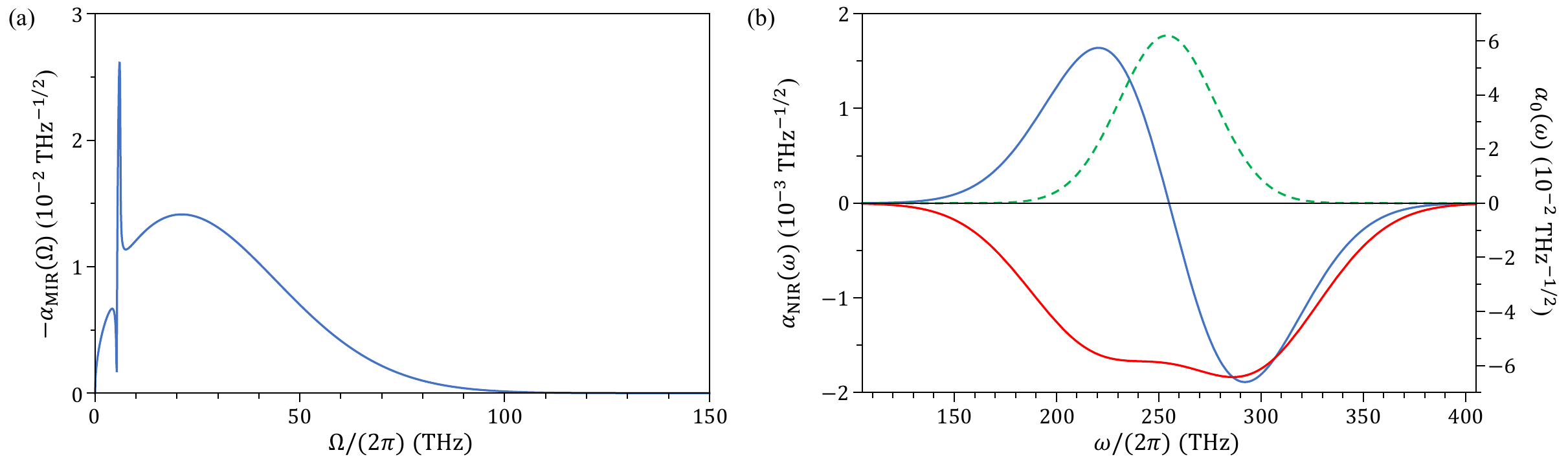"}
	\caption{
		\label{Fig:spectral amplitude of detection mode}
(a) Spectral amplitude of the contribution to the detection mode in the MIR frequency range $\alpha_{\mathrm{MIR}}(\Omega)=-\beta_{\mathrm{MIR}}(\Omega)$. Only the nonvanishing real part is shown.
(b)  Real (imaginary) part of the spectral amplitude of the contribution to the detection mode in the NIR frequency range $\alpha_{\mathrm{NIR}}(\omega)=-\beta_{\mathrm{NIR}}(\omega)$ in blue (red) color, while the spectral amplitude of the zeroth-order contribution $\alpha_{0}(\omega)$ is plotted as a green dashed line.
The probe pulse is Gaussian with central frequency $\omega_\mathrm{c}/(2\pi)=255\,$THz and bandwidth $\Delta \omega/(2\pi)=33\,$THz.
}
\end{figure}
To this end, we consider a probe pulse $E_{\mathrm{p},z}(t)=\mathcal{E} \sin(\omega_\mathrm{c} t) e^{-(t/\Delta t)^2}$ with a central frequency $\omega_\mathrm{c}/(2\pi)=255\,$THz and a temporal duration $\Delta t=9.65\,$fs, corresponding to a spectral bandwidth of $\Delta \omega/(2\pi)=33\,$THz.
For the EOX, we consider a $l=7\mathrm{\mu m}$ thick ZnTe with the RI $n=2.76$ and the group RI $n_\mathrm{g}=2.9$ for the NIR frequency.
For the MIR frequency range, the RI is computed using a Lorentz model
\begin{eqnarray}
	n(\Omega)=\Re\sqrt{6.7\bigg( 1+\frac{6.2^2-5.3^2}{5.3^2-\tilde{\Omega}^2-0.09 i \lvert \tilde{\Omega} \rvert } \bigg)},
\end{eqnarray}
where $\tilde{\Omega}=\Omega/(2\pi\, \mathrm{THz})$ \cite{Guedes2023,MoskalenkoParaxial}.
We choose the number of photons in the probe to be $N=2.3\times 10^{10}$ with the transverse area $A=\pi (3\mu \text{m})^2$, such that the effective interaction strength defined as $r=\lvert\frac{dl}{nc_0}\rvert \sqrt{\omega_\mathrm{c}\Delta\omega} E_0$ \cite{Wasilewski2006,KizmannSubcycle}, where $E_0=\lvert \int_0^\infty \mathrm{d}\omega E_{\mathrm{p},z}(\omega) \rvert $ and $d=-232\,\text{pm/V}$, amounts to $r=1$.
The spectral amplitudes $\alpha(\omega)$ of the
$\hat{\alpha}$ modes
are defined via
\begin{eqnarray}\label{eq:spectral amplitudes of alpha modes}
	\begin{split}
		\hat{\alpha}_{0}(t_\mathrm{d})=& \int_{\Omega_{\mathrm{max}}}^\infty \mathrm{d}\omega \, \alpha_{0}(\omega) \hat{a}(\omega) e^{-i\omega t_\mathrm{d}}
		,\\
		\hat{\alpha}_{\mathrm{MIR}}(t_\mathrm{d})=&\int_0^{\Omega_{\mathrm{max}}} \mathrm{d}\Omega \, \alpha_{\mathrm{MIR}}(\Omega) \hat{a}(\Omega)e^{-i\Omega t_\mathrm{d}}
		,\\
		\hat{\alpha}_{\mathrm{NIR}}(t_\mathrm{d})=&\int_{\Omega_{\mathrm{max}}}^\infty \mathrm{d}\omega \, \alpha_{\mathrm{NIR}}(\omega) \hat{a}(\omega) e^{-i\omega t_\mathrm{d}}.
	\end{split}
\end{eqnarray}
Spectral amplitudes of the $\hat{\beta}$ modes can also be defined in a similar manner.
Now, by utilizing Eqs.~\eqref{eq:THz detection mode},\eqref{eq:NIR alpha detection mode}, and \eqref{eq:NIR beta detection mode}, we derive the expressions for spectral amplitudes in the MIR range, i.e., $\alpha_{\mathrm{MIR}}(\Omega)$ and $\beta_{\mathrm{MIR}}(\Omega)$, as well as in the NIR range, i.e., $\alpha_{0}(\omega)$, $\alpha_{\mathrm{NIR}}(\omega)$ and $\beta_{\mathrm{NIR}}(\omega)$. The results
are shown in Fig.~\ref{Fig:spectral amplitude of detection mode}.
The magnitude of the detection mode for higher-order contributions becomes smaller, which validates the perturbation scheme and supports its interpretation as operating in a back-action free regime \cite{Guedes2023}: $[\hat{\alpha}_{0}(t_\mathrm{d}),\hat{\alpha}^\dagger_{0}(t_\mathrm{d})]=1$, $[\hat{\alpha}_{\mathrm{MIR}}(t_\mathrm{d}),\hat{\alpha}^\dagger_{\mathrm{MIR}}(t_\mathrm{d})]=4.6\times 10^{-2}$ and $[\hat{\alpha}_{\mathrm{NIR}}(t_\mathrm{d}),\hat{\alpha}^\dagger_{\mathrm{NIR}}(t_\mathrm{d})]=4.1\times 10^{-3}$.

Next, we analyze the signal operator in Eq.~\eqref{eq:field_oper} by decomposing it into MIR and NIR frequency components as follows: $\hat{S}(\theta;t_\mathrm{d})=\hat{S}_{\mathrm{MIR}}(\theta;t_\mathrm{d})+\hat{S}_{\mathrm{NIR}}(\theta;t_\mathrm{d})$, where  $\hat{S}_{\mathrm{MIR}}(\theta;t_\mathrm{d})=\sqrt{N}[P(\theta)(\hat{\alpha}_{\mathrm{MIR}}(t_\mathrm{d})+\hat{\beta}^\dagger_{\mathrm{MIR}}(t_\mathrm{d}))+\text{H.c.}]$ and $\hat{S}_{\mathrm{NIR}}(\theta;t_\mathrm{d})=\sqrt{N}[P(\theta)(\hat{\alpha}_{0}(t_\mathrm{d})+\hat{\alpha}_{\mathrm{NIR}}(t_\mathrm{d})+\hat{\beta}^\dagger_{\mathrm{NIR}}(t_\mathrm{d}))+\text{H.c.}]$.
By assuming the $s$-polarized NIR field (quantum MIR) to be in the vacuum state $\ket{0_\mathrm{NIR}}$ (state $\ket{\psi_{\mathrm{MIR}}}$) and making use of above decomposition, the variance of the signal can be approximately written as
 \begin{eqnarray}\label{eq:one-port EOS decomposition}
 	\begin{split}
 		\langle \hat{S}^2(\theta;t_\mathrm{d}) \rangle
 		&\approx N\bra{0_\mathrm{NIR}} \hat{\alpha}_{0}(t_\mathrm{d})\hat{\alpha}^\dagger_{0}(t_\mathrm{d}) \ket{0_\mathrm{NIR}}
 		\\
 		&+N\bra{\psi_{\mathrm{MIR}}} \big( P(\theta)\hat{\alpha}_{\mathrm{MIR}}(t_\mathrm{d})+P(\theta)\hat{\beta}^\dagger_{\mathrm{MIR}}(t_\mathrm{d})+\text{H.c.} \big)^2\ket{\psi_{\mathrm{MIR}}}
 		\\
 		&+N\bra{0_\mathrm{NIR}} P^2(\theta)\hat{\alpha}_{0}(t_\mathrm{d})\hat{\beta}^\dagger_{\mathrm{NIR}}(t_\mathrm{d}) +\hat{\alpha}_{0}(t_\mathrm{d})\hat{\alpha}^\dagger_{\mathrm{NIR}}(t_\mathrm{d})+\text{H.c.} \ket{0_\mathrm{NIR}}.
 	\end{split}
 \end{eqnarray}
Here, we considered terms up to the second order of the effective interaction strength.
The first term on the right-hand side (rhs) of Eq.~\eqref{eq:one-port EOS decomposition} corresponds to the shot noise contribution $N$.
The second term on the rhs represents the contribution to the variance of the signal originating from the MIR field $\langle \hat{S}^2_{\mathrm{MIR}}(\theta;t_\mathrm{d}) \rangle$.

For any NIR probe pluse $E_{\mathrm{p},z}(t)$, the annihilation operators $\hat{\alpha}_{\mathrm{MIR}} (t_\mathrm{d})$ and $\hat{\beta}_{\mathrm{MIR}} (t_\mathrm{d})$
are connected as $\hat{\alpha}_{\mathrm{MIR}}(t_\mathrm{d})=-\hat{\beta}_{\mathrm{MIR}}(t_\mathrm{d})$.
Consequently, the MIR signal in the EOS can be written as $\hat{S}_{\mathrm{MIR}}(\theta;t_\mathrm{d})=2\sqrt{2N}\sin \!\phi\, \hat{p}_{\mathrm{MIR}}(t_\mathrm{d})$, where we use $P(\theta)=e^{-i\phi}$ and $\phi=\arcsin(-\sqrt{2}\cos\frac{\theta}{2})$ with $-\pi/2\leq\phi \leq \pi/2$. The MIR quadrature is defined as $\hat{p}_{\mathrm{MIR}}(t_\mathrm{d})=\frac{i}{\sqrt{2}}[\hat{\alpha}^\dagger_{\mathrm{MIR}}(t_\mathrm{d})-\hat{\alpha}_{\mathrm{MIR}}(t_\mathrm{d})]$ and the NIR quadrature $\hat{p}_{\mathrm{NIR}}(t_\mathrm{d})$ is defined in a similar manner.
In result, the variance of the MIR signal is given by $\langle \hat{S}^2_{\mathrm{MIR}}(\theta;t_\mathrm{d}) \rangle= 8N \sin^2\!\phi\, \langle \hat{p}^2_{\mathrm{MIR}}(t_\mathrm{d}) \rangle$, which shows that the MIR contribution to the total EOS signal
corresponds only to
the $p$-quadrature of $\hat{\alpha}_{\mathrm{MIR}}(t_\mathrm{d})$, even with various phase shifts.

The third term on the rhs of Eq.~\eqref{eq:one-port EOS decomposition} represents the contribution to the variance of the signal coming from the NIR field, excluding the shot noise: $\langle \hat{S}^2_{\mathrm{NIR}}(\theta;t_\mathrm{d}) \rangle-N$.
It consists of two terms, the phase dependent (PD) term $\bra{0_\mathrm{NIR}} P^2(\theta)\hat{\alpha}_{0}(t_\mathrm{d}) \hat{\beta}^\dagger_{\mathrm{NIR}}(t_\mathrm{d}) +\text{H.c.} \ket{0_\mathrm{NIR}}$ and the phase independent (PI) term $\bra{0_\mathrm{NIR}} \hat{\alpha}_{0}(t_\mathrm{d})\hat{\alpha}^\dagger_{\mathrm{NIR}}(t_\mathrm{d})+\text{H.c.}\ket{0_\mathrm{NIR}}$.
Since the spectral amplitude of $\hat{\alpha}_{0}(t_\mathrm{d})$ and the real part of the spectral amplitude of $\hat{\alpha}_{\mathrm{NIR}}(t_\mathrm{d})$ are mutually orthogonal (cf. Fig.~\ref{Fig:spectral amplitude of detection mode}b)
, the contribution of the PI term to the signal variance vanishes. This means that
the contribution where each of the consecutive $\chi^{(2)}$ process conserves energy does not affect the variance, as shown in  \cite{Kizmann2022Theory,Kizmann2022}.
Further, only the imaginary part of $\beta_{\mathrm{NIR}}(\omega)$ in the PD term (cf. Fig.~\ref{Fig:spectral amplitude of detection mode}b) leads to a contribution, which can be expressed as $\bra{0_\mathrm{NIR}} P^2(\theta)\hat{\alpha}_{0}(t_\mathrm{d}) \hat{\beta}^\dagger_{\mathrm{NIR}}(t_\mathrm{d}) +\text{H.c.} \ket{0_\mathrm{NIR}}=2\sin(2\phi)\int_{\Omega_{\mathrm{max}}}^\infty \mathrm{d}\omega \, \alpha_0(\omega) \Im\alpha_{\mathrm{NIR}}(\omega)$ by using the relation $\alpha_{\mathrm{NIR}}(\omega)=-\beta_{\mathrm{NIR}}(\omega)$, which holds for any NIR probe pulse.
This indicates that only cascading processes, where energy is conserved when both $\chi^{(2)}$ sub-processes are considered together,
provide a nonvanishing contribution
\cite{Kizmann2022Theory,Kizmann2022}.
In an actual experiment, the MIR part
$\langle \hat{S}^2_{\mathrm{MIR}}(\theta;t_\mathrm{d}) \rangle$ can be identified by separating the NIR part
$\langle \hat{S}^2_{\mathrm{NIR}}(\theta;t_\mathrm{d}) \rangle$ from the measured signal variance $\langle \hat{S}^2(\theta;t_\mathrm{d}) \rangle$ \cite{Yang2023}:
  \begin{eqnarray}\label{eq:THz signal on one-port EOS}
 	\begin{split}
 		\langle \hat{S}^2_{\mathrm{MIR}}(\theta;t_\mathrm{d}) \rangle \;=\;
 \langle \hat{S}^2(\theta;t_\mathrm{d}) \rangle-N-2\sin(2\phi)\int_{\Omega_{\mathrm{max}}}^\infty \mathrm{d}\omega \, \alpha_0(\omega) \Im\alpha_{\mathrm{NIR}}(\omega).
 	\end{split}
 \end{eqnarray}

\section{two-port EOS}\label{sec:two-port EOS}

\begin{figure}[t]\includegraphics[width=18cm]{"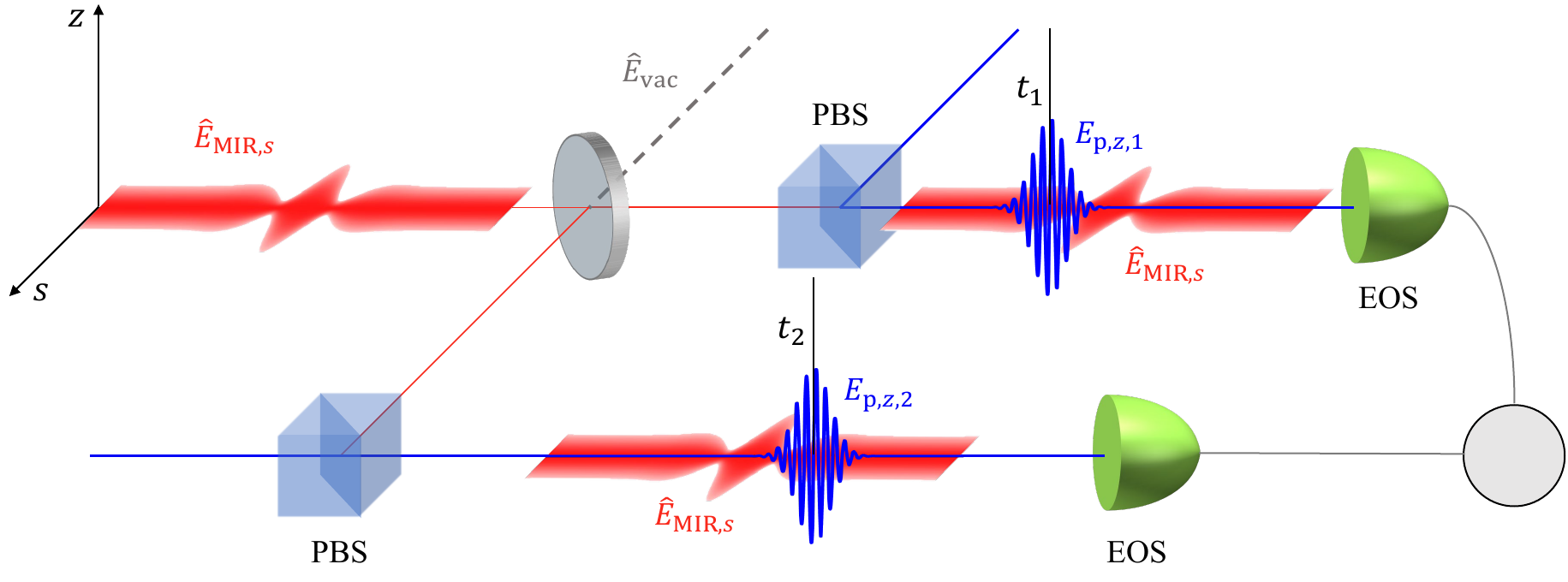"}
	\caption{
		\label{Fig:two-port EOS setup}
		Schematic of the two-port EOS setup to measure the temporal correlation. The MIR QLP $\hat{E}_{\mathrm{MIR},s}$ (red band) is divided into two parts by a balanced (50/50) beam splitter, where the complementary vacuum $\hat{E}_{\mathrm{vac}}$ (gray dashed line) coming from the idle port of the beam splitter is admixed. Polarizing beam splitters (PBSs) are used to impose the NIR probe pulses $E_{\mathrm{p},z,\mathrm{1}}$ and $E_{\mathrm{p},z,\mathrm{2}}$ (blue)
having polarization in the perpendicular direction with respect to the tested wave, with independently tunable time delays $t_1$ and $t_2$.
Signals of the two EOS channels are accumulated to retrieve the
joint probability distribution $P(S_1,S_2)$.}
\end{figure}

 In general, ultrafast QLPs to be characterized via the EOS occupy an ultrabroadband frequency range in the MIR and
can encompass multiple principal modes.
To extract these modes, reconstruction of the coherency matrix in the frequency domain is needed \cite{Fabre2020}. In the ultrabroadband case, its direct acquisition by the methods of conventional frequency-domain quantum optics is hindered by a huge number of required frequency modes and their minor overlap with the underlying strongly localized modes of the QLP. A reconstruction of this matrix happens to be also impossible based on the single-port EOS setup.
To overcome this limitation, we utilize a variation of the two-port EOS setup \cite{Benea2019}, as illustrated in Fig.~\ref{Fig:two-port EOS setup}, to construct a temporal coherence matrix, which will later serve as the basis for its frequency-domain counterpart.
In this setup, the ellipsometry embedded in the two EOS channels operates analogously to the homodyne detection \cite{MoskalenkoParaxial,Yang2023}.
So, the joint probability distribution of the two-port EOS signal detected via the ellipsometry can be expressed as
\begin{eqnarray}\label{eq:photodetection}
	P(S_1,S_2)=\Big\langle:\frac{1}{\sqrt{2\pi N_1}}e^{-\frac{(S_1-\hat{S}_1(\theta_1;t_1))^2}{2N_1}}\frac{1}{\sqrt{2\pi N_2}}e^{-\frac{(S_2-\hat{S}_2(\theta_2;t_2))^2}{2N_2}}:\Big\rangle,
\end{eqnarray}
where $\langle :\quad : \rangle$ denotes normally ordered expectation value, the single-channel signal operators are defined as $\hat{S}_j(\theta_j;t_j)=\sqrt{N_j}\Big(P(\theta_j)\hat{b}_j(t_j)+\mathrm{H.c.}\Big)$, the time-delayed NIR probe pulses for each port have the same shape $E_{\mathrm{p},z,j}(t)= E_{\mathrm{p},z}(t-t_j)$ up to the delay, and $N_j\equiv N=C\int_0^\infty \mathrm{d}\omega \, \frac{1}{\hbar \omega} \lvert E_{\mathrm{p},z}(\omega) \rvert^2$ with $j\in \{1,2\}$ \cite{Kizmann2022,Vogel2006}.
The corresponding detection mode operators for each EOS channel can be expressed in a similar way as in
Eq.~\eqref{eq:field_oper_2} with $t_\mathrm{d}=t_j$,
\begin{eqnarray}\label{eq:twoportbmode}
	\hat{b}_j(t_j)=\int_0^\infty \mathrm{d} \omega \, h(\omega;t_j) \hat{b}_j(\omega),
	\quad
	h(\omega;t_j)=i\sqrt{\frac{C}{\hbar N}} \frac{E_{\mathrm{p},z}^*(\omega)e^{-i\omega t_j}}{\sqrt{\omega}}.
\end{eqnarray}
Next, by rescaling parameters $x_j = \frac{S_j}{\sqrt{2N}}$ and defining quadrature operators $\hat{x}_j(\phi_j;t_j) = \frac{1}{\sqrt{2}}\big(e^{-i\phi_j}\hat{b}_j(t_j) + \mathrm{H.c.}\big)$, we can rewrite the joint probability distribution of the signal as:
\begin{equation}\label{eq:photodetectionquadrature}
	P(x_1, x_2) = \Big\langle : \frac{1}{2\pi N}
	\exp\big[-(x_1 - \hat{x}_1(\phi_1;t_1))^2
	- (x_2 - \hat{x}_2(\phi_2;t_2))^2\big] : \Big\rangle .
\end{equation}
The above joint probability distribution can be simplified in the Fourier space, introducing the characteristic function
as $\widetilde{P}(u,v)=\iint \mathrm{d}x_1 \mathrm{d}x_2 \, P(x_1,x_2) e^{-i(x_1 u+x_2 v)}$, leading to
\begin{eqnarray}\label{eq:photodetectioncharacteristic}
	\widetilde{P}(u,v)
	=\frac{1}{2N} \big\langle e^{-iu\hat{x}_1(\phi_1;t_1)-iv\hat{x}_2(\phi_2;t_2)} \big\rangle .
\end{eqnarray}
Now the statistical moments of the signal can be related to the mean values of the quadratures by the following expression \cite{LeonhardtEssential}:
\begin{equation}\label{eq:Weyl correspondence}
	\begin{split}
		i^n\frac{\partial}{\partial \xi^n} & \tilde{P}(\xi \lambda,\xi \mu)\Big\vert_{\xi=0}=\iint \mathrm{d}x_1  \mathrm{d}x_2\, (\lambda x_1+\mu x_2)^n P(x_1,x_2)
		\\
		&=\frac{1}{2N} \big\langle (\lambda \hat{x}_1(\phi_1;t_1)+\mu \hat{x}_2(\phi_2;t_2))^n \big\rangle .
	\end{split}
\end{equation}
Utilizing the above equation, we can extract the second moments by comparing the coefficients of $\lambda$ and $\mu$, and write them as
\begin{equation}\label{eq:Weyl correspondence}
	\begin{split}
		\iint \mathrm{d}x_1  \mathrm{d}x_2\, x_1^2 P(x_1,x_2)&=\frac{1}{2N}\langle \hat{x}^2_1(\phi_1;t_1)\rangle,
		\\
		\iint \mathrm{d}x_1  \mathrm{d}x_2\, x_2^2 P(x_1,x_2)&=\frac{1}{2N}\langle \hat{x}^2_2(\phi_2;t_2)\rangle,
		\\
		\iint \mathrm{d}x_1  \mathrm{d}x_2\, x_1 x_2 P(x_1,x_2)&=\frac{1}{2N} \Big\langle \frac{1}{2} \big\{\hat{x}_1(\phi_1;t_1),\hat{x}_2(\phi_2;t_2) \big\} \Big\rangle .
		\end{split}
\end{equation}
Following a similar approach as in Sec.~\ref{sec:One-port EOS}, the quadrature corresponding to each EOS channel can be expressed in terms of the quadratures of the MIR field, NIR field, and complementary vacuum field, originating from the idle port of the first beam splitter (see Fig.~\ref{Fig:two-port EOS setup}), as
\begin{equation}\label{eq:input quadrature}
	\begin{split}
		\hat{x}_j(\phi_j;t_j)=&\frac{1}{\sqrt{2}} \big(2\sin\phi_j \hat{p}_{\mathrm{MIR}}(t_j)\big)
		\\ + &\frac{1}{\sqrt{2}} \big(\hat{x}_0(\phi_j;t_j)+2\sin\phi_j \hat{p}_{\mathrm{NIR}}(t_j) \big)
		\\ + &\frac{(-1)^{j+1}}{\sqrt{2}} \big( \hat{x}_{0,\mathrm{vac}}(\phi_j;t_j)+  2\sin\phi_j \hat{p}_{\mathrm{MIR,vac}}(t_j)+2\sin\phi_j \hat{p}_{\mathrm{NIR,vac}}(t_j) \big),
	\end{split}
\end{equation}
where the zeroth-order quadrature is defined as $\hat{x}_0(\phi_j;t_j)=\frac{1}{\sqrt{2}}\big(e^{-i\phi_j}\hat{\alpha}_0(t_j) + \mathrm{H.c.}\big)$, $\hat{p}_{\mathrm{MIR}}$ and $\hat{p}_{\mathrm{NIR}}$ are defined in the end of Sec.~\ref{sec:One-port EOS}, and the subscript ``$\mathrm{vac}$'' denotes the corresponding operator
belonging to the complementary vacuum field.
Here, we considered the transformation of the incoming fields imposed by the balanced beam splitter and used the relations $\hat{\beta}_{\mathrm{MIR}}(t_j)=-\hat{\alpha}_{\mathrm{MIR}}(t_j)$ and $\hat{\beta}_{\mathrm{NIR}}(t_j)=-\hat{\alpha}_{\mathrm{NIR}}(t_j)$, which are valid as discussed in Sec.~\ref{sec:One-port EOS}.
Now by making use of Eq.~\eqref{eq:input quadrature}, the covariance of quadratures representing the EOS signals can be expressed as
\begin{equation}\label{eq:correlation}
	\begin{split}
		\Big\langle \frac{1}{2} \big\{\hat{x}_1 (\phi_1;t_1),\hat{x}_2 (\phi_2;t_2) \big\} \Big\rangle=
		\sin\phi_1 \sin\phi_2 \big[\bra{\psi_\mathrm{MIR}} \big\{ \hat{p}_{\mathrm{MIR}}(t_1), \hat{p}_{\mathrm{MIR}}(t_2) \big\}  \ket{\psi_\mathrm{MIR}} &
		\\
		- \bra{0} \big\{ \hat{p}_{\mathrm{MIR,vac}}(t_1), \hat{p}_{\mathrm{MIR,vac}}(t_2) \big\}  \ket{0} & \big].
	\end{split}
\end{equation}
The correlation of the NIR component is canceled by the corresponding NIR component of the complementary vacuum. Although we assume a balanced beam splitter for simplicity, this cancellation originates from the Stokes relations and therefore holds for any beam splitter.

\begin{sloppypar}
Taking into account
Eq.~\eqref{eq:Weyl correspondence} for the left hand side of Eq.~\eqref{eq:correlation},
we can express the correlation of the MIR frequency component as
\begin{eqnarray}\label{eq:anticommutation of THz parts}
	\begin{split}
	 \bra{\psi_\mathrm{MIR}} \big\{ \hat{p}_{\mathrm{MIR}}(t_1), \hat{p}_{\mathrm{MIR}}(t_2) \big\}  \ket{\psi_\mathrm{MIR}}=&
		\frac{2N}{\sin\phi_1 \sin\phi_2}\bigg(\iint \mathrm{d}x_1  \mathrm{d}x_2\, x_1 x_2 P(x_1,x_2) \bigg)
		\\
		&+\bra{0} \big\{ \hat{p}_{\mathrm{MIR}}(t_1), \hat{p}_{\mathrm{MIR}}(t_2) \big\}  \ket{0}.
	\end{split}
\end{eqnarray}
Using the identity $\bra{\psi_\mathrm{MIR}} \big[ \hat{p}_{\mathrm{MIR}}(t_1), \hat{p}_{\mathrm{MIR}}(t_2) \big] \ket{\psi_\mathrm{MIR}}=\bra{0} \big[ \hat{p}_{\mathrm{MIR}}(t_1), \hat{p}_{\mathrm{MIR}}(t_2) \big]  \ket{0}$ and evaluating $\bra{0} \hat{p}_{\mathrm{MIR}}(t_1)   \hat{p}_{\mathrm{MIR}}(t_2)  \ket{0}$,
we can then write the temporal coherence matrix $C(t_1,t_2)=\bra{\psi_\mathrm{MIR}} \hat{p}_{\mathrm{MIR}}(t_1)   \hat{p}_{\mathrm{MIR}}(t_2)  \ket{\psi_\mathrm{MIR}}$
as
\end{sloppypar}
\begin{eqnarray}\label{eq:correlation through vacuum}
		C(t_1,t_2)=
\frac{N}{\sin\phi_1 \sin\phi_2}\bigg(\iint \mathrm{d}x_1  \mathrm{d}x_2\, x_1 x_2 P(x_1,x_2) \bigg)
		+\frac{1}{2} \int_0^{\Omega_{\mathrm{max}}} \mathrm{d}\Omega \, \lvert \alpha_{\mathrm{MIR}}(\Omega) \rvert ^2 e^{i \Omega (t_2-t_1)}.
\end{eqnarray}
It is interesting to point out that in contrast to the one-port EOS [cf. Eq.~\eqref{eq:THz signal on one-port EOS}],  the derivation of the second term on the rhs of Eq.~\eqref{eq:correlation through vacuum} ultimately requires only first-order perturbation contributions, with no contributions from the NIR signal, effectively eliminating effects of the shot noise and cascading processes.
The first term on the rhs of Eq.~\eqref{eq:correlation through vacuum} can be acquired experimentally. Even if the phases $\phi_1$ and $\phi_2$ appear in both the prefactor and the joint probability distribution of the signal $P$, this term is finally independent of these phases.
Here there is some freedom concerning experimental implementation and its verification in terms of consistency.
The second term on the rhs of Eq.~\eqref{eq:correlation through vacuum} has to be calculated theoretically, taking into account the information on the probe pulse and parameters of the nonlinear interaction, contained, e.g., in Eqs.~\eqref{eq:F1},\eqref{eq:THz detection mode},\eqref{eq:Gpm} and Fig.~\ref{Fig:spectral amplitude of detection mode}a.
In result, complete information about the temporal coherence matrix of the MIR quantum light can be obtained.

\section{Reconstruction of Gaussian QLPs}\label{sec:Reconstruction of Gaussian quantum light pulse}
In the previous section, we explained how to get the temporal coherence matrix of a MIR Gaussian QLP.
Once the coherence matrix is obtained, it can be used to extract the principal modes of this pulsed light
and the second moments of the quadratures belonging to these modes,
thereby enabling its complete characterization.
In this section, we outline the corresponding procedure.
At first, we use the definition $\hat{p}_{\mathrm{MIR}}(t)=i(\hat{\alpha}^\dagger_{\mathrm{MIR}}(t)-\hat{\alpha}_{\mathrm{MIR}}(t))/\sqrt{2}$ and rewrite the expression for the temporal coherence matrix via its frequency domain components as
\begin{eqnarray}\label{eq:temporal correlations on THz frequency domain}
		C(t_1,t_2)=&\frac{1}{2} \int_0^{\Omega_\mathrm{max}}\hspace{-0.7cm}\mathrm{d} \Omega_1 \int_0^{\Omega_\mathrm{max}}\hspace{-0.7cm}\mathrm{d} \Omega_2 \,
\sum_{n,m=0}^1 (-1)^{1+n+m} \hat{\mathcal{C}}_1^n\hat{\mathcal{C}}_2^m
\alpha_{\mathrm{MIR}}(\Omega_1)  \alpha_{\mathrm{MIR}}(\Omega_2) e^{-i \Omega_1 t_1 - i \Omega_2 t_2} \langle  \hat{a}(\Omega_1) \hat{a}(\Omega_2) \rangle,
\end{eqnarray}
where $\hat{\mathcal{C}_1}$ ($\hat{\mathcal{C}_2}$) reflects Hermitian conjugation of the $\Omega_1$ ($\Omega_2$) dependent part.
In the above representation, the annihilation and creation operators contribute to the positive and negative frequency parts of $C(t_1,t_2)$, respectively.
Then, by performing a two-dimensional Fourier transform of Eq.~\eqref{eq:temporal correlations on THz frequency domain}, we
find
\begin{eqnarray}\label{eq:correlations on THz frequency}
	\langle  \hat{a}(\Omega_1) \hat{a}(\Omega_2) \rangle =-\frac{2}{(2\pi)^2}\frac{1}{\alpha_{\mathrm{MIR}}(\Omega_1) \alpha_{\mathrm{MIR}}(\Omega_2)}\int \mathrm{d}t_1 \int \mathrm{d}t_2 \, C(t_1,t_2) e^{i \Omega_1 t_1 +i \Omega_2 t_2}.
\end{eqnarray}
Knowing the mean values on the lhs of Eq.~\eqref{eq:correlations on THz frequency}
allows to determine any second moments of field operators in the MIR range, confined by the condition that $\alpha_{\mathrm{MIR}}(\Omega)$ does not become negligibly small.
Furthermore, the spectral profile of $\alpha_{\mathrm{MIR}}(\Omega)$ can be experimentally determined by substituting the MIR quantum light with a reference coherent field having well-characterized spectral amplitudes $\alpha_\mathrm{ref}(\Omega)$.
For instance, by evaluating the Fourier components of the temporal coherence matrix for a negative frequency $\Omega_1=-\Omega$ and a positive frequency $\Omega_2=\Omega$ using Eq.~\eqref{eq:correlation through vacuum},
we can obtain
\begin{eqnarray}\label{eq:spectrum of alpha MIR}
	\lvert\alpha_{\mathrm{MIR}}(\Omega)\rvert^2=\frac{2\lvert\alpha_\mathrm{ref}(\Omega)\rvert^2}{(2\pi)^2} \int \mathrm{d}t_1 \int \mathrm{d}t_2 \, \frac{N}{\sin\phi_1 \sin\phi_2}\bigg(\iint \mathrm{d}x_1  \mathrm{d}x_2\, x_1 x_2 P_\mathrm{ref}(x_1,x_2) \bigg) e^{i \Omega (t_2-t_1)}.
\end{eqnarray}
Since $\alpha_{\mathrm{MIR}}(\Omega)$ is a real function, we can get it now from Eq.~\eqref{eq:spectrum of alpha MIR}, up to the sign. However, the sign does not matter for both $\lvert\alpha_{\mathrm{MIR}}(\Omega)\rvert^2$ and $\alpha_{\mathrm{MIR}}(\Omega_1)\alpha_{\mathrm{MIR}}(\Omega_2)$, used in Eqs.~\eqref{eq:correlation through vacuum} and \eqref{eq:correlations on THz frequency}, if we fix the sign convention. With this, eventually, all necessary information for these equations can be acquired experimentally.

In Sec.~\ref{sec:One-port EOS}, considering actual experimental parameters, we show in Fig.~\ref{Fig:spectral amplitude of detection mode}a that the frequency of MIR mode $\Omega$ ranges from $0$ to $\Omega_{\mathrm{max}}\approx 100$~THz.
This implies
that any multimode Gaussian QLP within this MIR frequency range can be fully characterized. We now elaborate the reconstruction process as follows: First, the principal modes are extracted from the eigenvectors of $\langle  \hat{a}^\dagger(\Omega_1) \hat{a}(\Omega_2) \rangle$ and express their operators as $\hat{a}_j=\int_0^{\Omega_{\mathrm{max}}} \mathrm{d}\Omega \, f_j(\Omega) \hat{a}(\Omega)$. Then, we evaluate the mean values of the second moments for these principal modes as
\begin{eqnarray}\label{eq:x variances of principal modes}
	\langle  \hat{x}_j^2 \rangle = \frac{1}{2} \int_0^{\Omega_{\mathrm{max}}}\hspace{-0.7cm} \mathrm{d} \Omega_1 \int_0^{\Omega_{\mathrm{max}}}\hspace{-0.7cm} \mathrm{d} \Omega_2 \,
\sum_{n,m=0}^1 \hat{\mathcal{C}}_1^n\hat{\mathcal{C}}_2^m
 f_j(\Omega_1) f_j(\Omega_2) \langle \hat{a}(\Omega_1) \hat{a}(\Omega_2)  \rangle,
\end{eqnarray}
\begin{eqnarray}\label{eq:p variances of principal modes}
	\langle  \hat{p}_j^2 \rangle = \frac{1}{2} \int_0^{\Omega_{\mathrm{max}}}\hspace{-0.7cm} \mathrm{d} \Omega_1 \int_0^{\Omega_{\mathrm{max}}}\hspace{-0.7cm} \mathrm{d} \Omega_2 \,
\sum_{n,m=0}^1 (-1)^{1+n+m} \hat{\mathcal{C}}_1^n\hat{\mathcal{C}}_2^m
 f_j(\Omega_1) f_j(\Omega_2) \langle \hat{a}(\Omega_1) \hat{a}(\Omega_2)  \rangle,
\end{eqnarray}
\begin{eqnarray}\label{eq:covariances of principal modes}
	 \langle  \{\hat{x}_j,\hat{p}_j\} \rangle = i \int_0^{\Omega_{\mathrm{max}}} \hspace{-0.7cm} \mathrm{d} \Omega_1 \int_0^{\Omega_{\mathrm{max}}} \hspace{-0.7cm} \mathrm{d} \Omega_2 \, \sum_{n=0}^1 (-1)^{1+n} \hat{\mathcal{C}}_1^n\hat{\mathcal{C}}_2^n f_j(\Omega_1) f_j(\Omega_2) \langle \hat{a}(\Omega_1) \hat{a}(\Omega_2)  \rangle,
\end{eqnarray}
where quadratures of the principal modes are defined by $\hat{x}_j=\frac{1}{\sqrt{2}}\big(\hat{a}_j+\hat{a}^\dagger_j \big)$ and $\hat{p}_j=\frac{i}{\sqrt{2}}\big(\hat{a}^\dagger_j-\hat{a}_j \big)$.
The second moments, represented by Eqs.~\eqref{eq:x variances of principal modes}-\eqref{eq:covariances of principal modes} can then be calculated from the reconstructed principal modes $f_j$ and Eq.~\eqref{eq:correlations on THz frequency}.

\begin{figure}[t]\includegraphics[width=18cm]{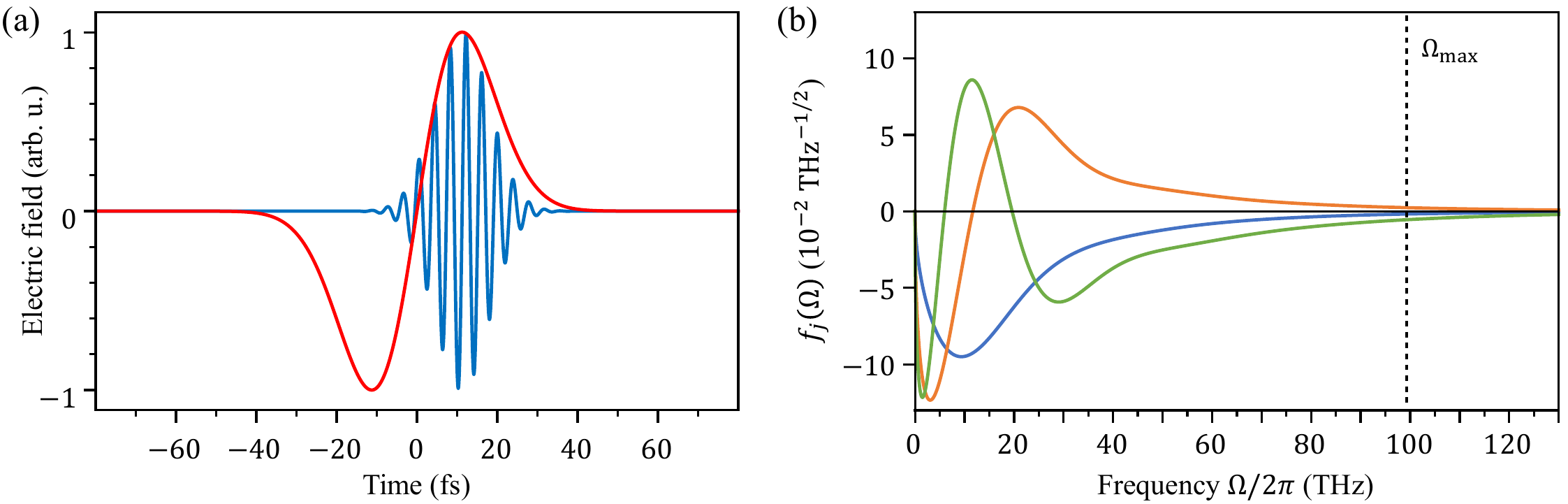}
	\caption{
		\label{Fig:squeezed state with single cycle driving}
		(a)
Single-cycle driving field, $E_{\mathrm{d}}=\mathcal{E}_{\mathrm{d}}\Gamma t e^{-(\Gamma t)^2}$, used for the generation of the multimode MIR QLP (red) and probe pulse used in the EOS (blue). (b) Spectral amplitudes of the dominant principal modes corresponding to generating QLP: First (blue), second (orange), and third (green) modes are shown. $\Omega_{\mathrm{max}}$ used in the numerical calculations is defined in Eq.~\eqref{eq:THz detection mode}. Parameters of the driving field are $\Gamma/(2\pi)=10$~THz and $r_\mathrm{g}=1$ \cite{KizmannSubcycle}, whereas the probe pulse has the same parameters as in Fig.~\ref{Fig:spectral amplitude of detection mode}.}
\end{figure}

\begin{figure}[t]\includegraphics[width=18cm]{"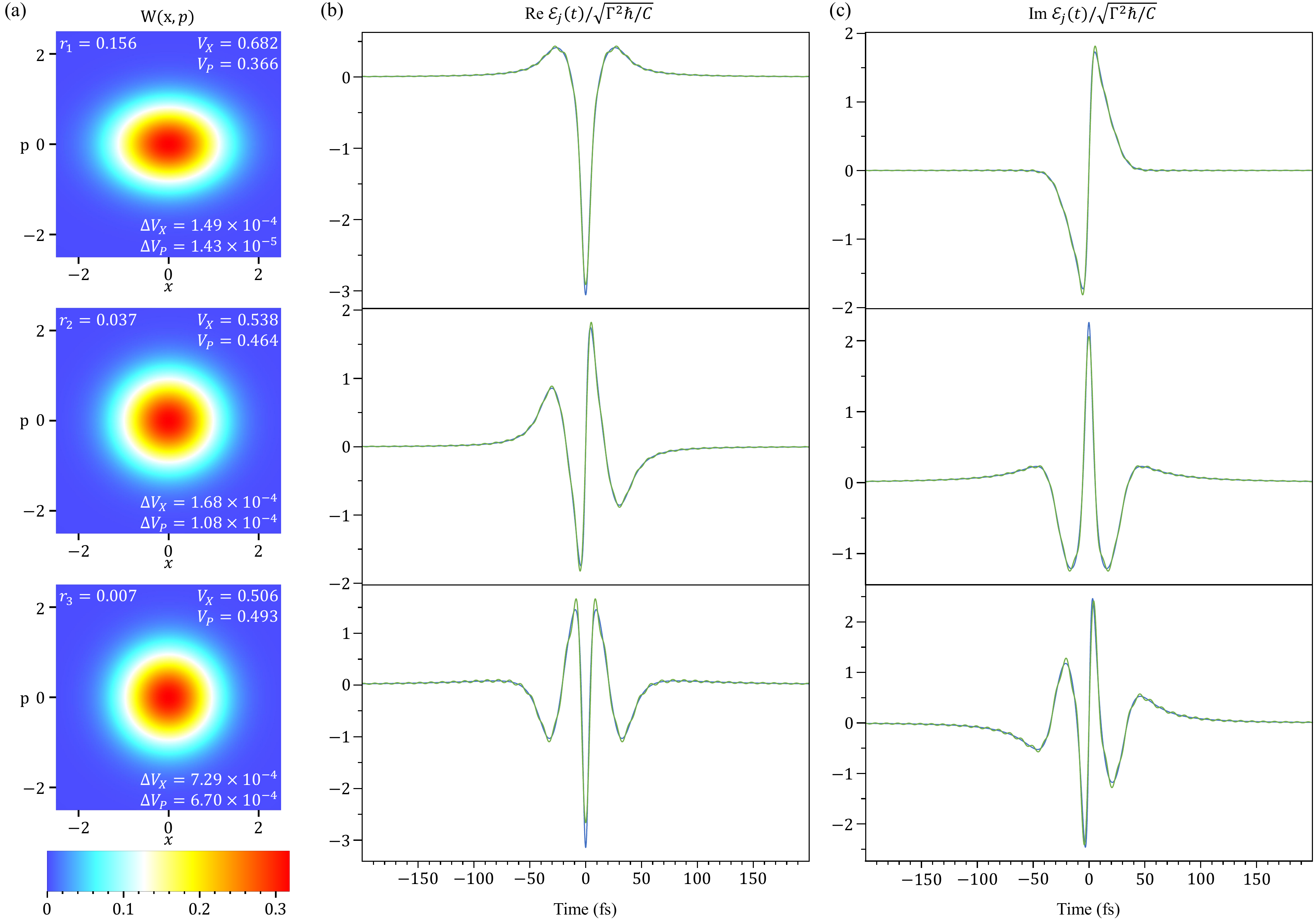"}
	\caption{
		\label{Fig:principal mode reconstruction}
	 Reconstruction of
the QLP corresponding to
Fig.~\ref{Fig:squeezed state with single cycle driving}.
		(a) Reconstructed Wigner functions of each principal mode up to the third (j=3) order. The respective actual squeezing factors $r_j$, reconstructed variances $V_X$ ($V_P$) for the $x$- ($p$-) quadrature as well as their deviations $\Delta V_X$ ($\Delta V_P$) from the actual variances are also indicated in the plots.
		The corresponding field modes $\mathcal{E}_j (t)=[\hat{E}^{(+)}(t),\hat{a}_j^\dagger]$ are arranged horizontally, with real parts shown in (b) and imaginary parts shown in (c). The exact wave forms are plotted in blue, whereas the reconstructed wave forms are plotted in green.
}
\end{figure}

We now use the outlined formalism
to reconstruct a Gaussian QLP using actual experimental parameters. Towards this, we consider an
ultrafast MIR Gaussian QLP generated via $\chi^{(2)}$ nonlinear interaction process driven by an exemplary single-cycle field $E_{\mathrm{d}}=\mathcal{E}_{\mathrm{d}}\Gamma t e^{-(\Gamma t)^2}$ with $\Gamma/(2\pi)=10$~THz.
We select the probe pulse for the EOS such that its duration is subcycle with respect to
the driving field, as shown in Fig.~\ref{Fig:squeezed state with single cycle driving}a.
A sufficiently thin generation crystal is considered to ensure a perfect phase matching.
The nonlinear interaction within the crystal comprises both sum-frequency and difference-frequency generation processes, leading to the effective interaction strength $r_\mathrm{g}=1$ \cite{KizmannSubcycle}.
Based on these parameters, we compute the spectral amplitudes $f_j(\Omega)$ of the principal modes for the generated ultrafast MIR QLP, which are shown in Fig.~\ref{Fig:squeezed state with single cycle driving}b.
Here, the phase degree of freedom of the principal modes is selected such that
the spectral amplitude $f_j(\Omega)$ becomes a real function, which leads to the variance of the $p$ ($x$) quadrature being maximized for the first and third (second) modes.
Next, by considering the detection mode $\alpha_{\mathrm{MIR}}(\Omega)$ as shown in Fig.~\ref{Fig:spectral amplitude of detection mode}a, we can reconstruct the principal modes of the MIR QLP based on the outlined two-port EOS technique.
The reconstructed principal modes are then compared with their exact counterparts, represented by the associated electric fields $\mathcal{E}_j(t)=[\hat{E}^{(+)}(t),\hat{a}_j]$, as depicted in Fig.~\ref{Fig:principal mode reconstruction}b and ~\ref{Fig:principal mode reconstruction}c.
These fields are related to the corresponding spectral amplitudes via $\mathcal{E}_j(t)=i\int_0^{\Omega_{\mathrm{max}}} \mathrm{d}\Omega \, \sqrt{\frac{\hbar \Omega}{C}} f^*_j(\Omega) e^{-i\Omega t}$, where the integrand includes a frequency-dependent part $\sqrt{\Omega}$ reflecting the ultrabroadband nature of the QLP.
The corresponding quantum states, illustrated by their Wigner functions in Fig.~\ref{Fig:principal mode reconstruction}a, are also reconstructed from $\langle \hat{x}_j^2 \rangle$ and $\langle \hat{p}_j^2 \rangle$, which are obtained from Eqs.~\eqref{eq:x variances of principal modes} and \eqref{eq:p variances of principal modes}, respectively.

The quantum electric field operator can be decomposed into a coherent amplitude and its purely quantum part, which captures the non-classical deviations from the mean field.
For a pure Gaussian QLP, based on the Bloch-Messiah reduction \cite{WeedbrookGaussian}, this decomposition can be explicitly expressed as
\begin{eqnarray}\label{eq:electric field reconstruction}
	 \hat{E}(t)=\sum_j \mathcal{E}_j(t) \, [\cosh r_j \, \hat{a}_j + e^{i\psi_j} \sinh r_j \, \hat{a}^\dagger_j+\alpha_j]+\text{H.c.},
\end{eqnarray}
where $\hat{a}_j$ are annihilation operators for the $j$-th principal modes in the vacuum picture \cite{Knight_book} and $\alpha_j$ denote the corresponding coherent amplitudes.
In the example shown in Fig.~\ref{Fig:principal mode reconstruction}, the phase factors of the temporal modes $\mathcal{E}_j(t)$ are chosen such that $\psi_j=0$, corresponding to squeezing along the $\hat{p}_j$ direction. The squeezing parameters can be evaluated from the reconstructed quadrature variances as $r_j=\frac{1}{2}\ln( 2 \langle \hat{x}_j^2 \rangle)=-\frac{1}{2}\ln(2  \langle \hat{p}_j^2 \rangle)$. Without loss of generality, we consider the case where the coherent amplitudes $\alpha_j$ vanish.

\section{Discussion}\label{sec:Discussion}

We demonstrated how the one-port EOS can be used to determine the detection modes in both the MIR and NIR frequency ranges.
This method enables the description of the EOS signal for an arbitrary incoming quantum state.
In particular, this study put focus on quantum states of the form $\ket{\psi_{\mathrm{MIR}}} \otimes \ket{0_\mathrm{NIR}}$.
For the special case of $\ket{0_{\mathrm{MIR}}} \otimes \ket{0_\mathrm{NIR}}$, decomposition of the one-port EOS signal as in Eqs.~\eqref{eq:one-port EOS decomposition} and \eqref{eq:THz signal on one-port EOS} reproduces the results of the previous studies \cite{MoskalenkoParaxial,Kizmann2022,Kizmann2022Theory}.
The MIR contribution to the signal variance, $\langle \hat{S}^2_{\mathrm{MIR}}(\theta;t_\mathrm{d}) \rangle$, corresponds to the ``classical'' term $\Gamma_\RN{1}$ defined in Ref.~\cite{Kizmann2022,Kizmann2022Theory}.
For the specific phase shift $\theta=\frac{\pi}{2}$, it represents $\langle  \hat{S}^2_\mathrm{EO} \rangle$ defined in Ref.~\cite{MoskalenkoParaxial}.
Similarly, subtraction of the shot noise contribution from the NIR part, $\langle \hat{S}^2_{\mathrm{NIR}}(\theta;t_\mathrm{d}) \rangle-N$, leads to the ``cascading'' term $\Gamma_\RN{3}$ defined in Ref.~\cite{Kizmann2022,Kizmann2022Theory}.

Reconstructing a quantum state typically involves accessing its phase-space distribution.
For the single-port EOS, it was found \cite{Sulzer2020,Yang2023,Onoe2023,Sinan2023,Emanuel2024} that spectral filtering is required to be employed to measure the phase-shifted quadratures of the MIR field.
For instance, it was theoretically demonstrated that it is possible to reconstruct an ultrafast single-mode QLP
using a spectral filter passing only photons from the low frequency tail of the probe spectrum to the balanced detectors
\cite{Yang2023}.
In contrast, in this work
we show that even a multimode Gaussian QLP
can be fully reconstructed using the two-port EOS, without the need for spectral filtering.
For that, it is important to have
the duration of the probe pulse
to be subcycle with respect to the driving field generating the QLP,
which is an inherent characteristic of EOS-based methods.
Consequently, the spectral amplitudes of the principal modes predominantly lie within the range of the MIR detection mode $\alpha_{\mathrm{MIR}}(\Omega)$ (see Fig.~\ref{Fig:spectral amplitude of detection mode}a), so that it is possible to select $\Omega_{\mathrm{max}}$ appropriately. The appropriate choice means that, on one hand side, the principal modes
are also sufficiently confined to the spectral interval bounded
by $\Omega_{\mathrm{max}}$, and, on the other hand side, the absolute value of $\alpha_{\mathrm{MIR}}(\Omega_\mathrm{max})$ does not become too small. This ensures the possibility of
successful and reliable reconstruction of the QLP based on the two-port EOS, as we described. Notice that in case of conventional frequency-domain approaches to quantum optics, which do not provide subcycle resolution, this possibility is obscured when sampling an ultrabroadband QLP since without the condition on the probe duration being fulfilled the detection mode $\alpha_{\mathrm{MIR}}(\Omega)$ would effectively vanish for a range of $\Omega$ still belonging to the spectral content of the QLP.
The oscillations observed in the reconstructed principal modes (cf. Figs.~\ref{Fig:principal mode reconstruction}b and ~\ref{Fig:principal mode reconstruction}c) arise from frequency components above $\Omega_{\mathrm{max}}$ in the generated quantum light.
These oscillations result mostly from the high-order cascading processes included in the theoretical model for generation of quantum light \cite{KizmannSubcycle}.
Further, a more careful consideration is required near $5.3$~THz, corresponding to the absorption peak of the EOX, as the smaller $\alpha_{\mathrm{MIR}}(\Omega)$ values, observed in Fig.~\ref{Fig:spectral amplitude of detection mode}a, may require enhanced detection precision when using Eq.~\eqref{eq:correlations on THz frequency}.

If the frequency range of QLP is known \textit{a priori}, two strategies can enhance the reconstruction accuracy.
First, the spectral bandwidth of the probe $\Delta \omega$ can be adjusted to align $\Omega_{\mathrm{max}}$ with the maximum frequency of the quantum light such that
vacuum contributions from irrelevant spectral regions can suppressed.
Second, the probe pulse shape can be optimized to obtain $\alpha_{\mathrm{MIR}}(\Omega)$ being the most efficient for the detection \cite{Onoe2023}.
While our protocol demonstrates
ability to recover
the principal modes of a Gaussian QLP, by the same method these modes can be obtained also in more general cases with the intrinsic separability of the quantum light \cite{Fabre2020,Yang2023}. However, this would not provide the complete information about such a non-Gaussian QLP in terms of its state or electric-field operator, depending if the Schr{\"{o}}dinger or Heisenberg picture is used for the description.
Future works might explore possibilities to extend the protocol towards a complete reconstruction of most general non-Gaussian QLPs.

Notably, similar to our variation of the two-port EOS setup, the original two-beam EOS configuration in Ref.~\cite{Lindel2024,Benea2019,Lindel2020} can also capture correlations of the THz electric field.
However, that two-beam EOS scheme may encounter practical challenges when probing correlations of electric field across the full MIR frequency range considered here, as ensuring sufficient spatial overlap at the detection plane becomes increasingly demanding at higher frequencies, introducing geometric constraints that are expected to influence the measurement accuracy \cite{Guedes2023}.
Moreover, this approach inherently includes the influence of MIR vacuum fields propagating in directions unrelated to the desired quantum signal.
In contrast, our version of the two-port EOS setup mitigates this issue by spatially separating the EOXs for both detection channels.

\section{Conclusion}\label{sec:conclusion}

In conclusion, we have investigated the ability of the single-port EOS and two-port EOS techniques to fully reconstruct the multimode structure of Gaussian QLPs in the MIR frequency range, ultimately enabling sampling of the corresponding electric-field operators.
Whereas, the single-port EOS fundamentally suffers from both the shot noise and cascading processes upon detection and enables a complete reconstruction only for pure Gaussian QLPs comprising just a single principal mode, we show that the two-port EOS technique can overcome these limitations.
In contrast to the previous studies that considered either local properties of the field or its various second-order correlations, our findings uplift EOS, focussing on tracing classical electric-field waveforms, to its complete quantum version, focussing on tracing electric-field operators for the generated quantum fields. This opens new avenues for time-domain quantum spectroscopies and metrological applications based on quantum light pulses.


\providecommand{\noopsort}[1]{}\providecommand{\singleletter}[1]{#1}%



\end{document}